\documentclass[prd,amsmath,amssymb,superscriptaddress,nofootinbib]{revtex4}
\usepackage{amsfonts}
\usepackage{multirow}
\usepackage{mathrsfs}
\usepackage{graphicx}
\usepackage{amsmath}
\usepackage{amssymb}
\usepackage{bm}
\usepackage{color}
\usepackage{diagbox}
\usepackage{hyperref}
\usepackage{slashed}

\usepackage{ulem} 

\newcommand{\nn}{\nonumber}

\newcommand{\beq}{\begin{equation}}
\newcommand{\eeq}{\end{equation}}
\newcommand{\bqa}{\begin{eqnarray}}
\newcommand{\eqa}{\end{eqnarray}}

\newcommand{\bseq}{\begin{subequations}}
\newcommand{\eseq}{\end{subequations}}

\def\sll{s_{ll}}

\makeatletter

\begin{document}

\title{Vector meson production associated with a lepton pair in $e^+$ $e^-$ annihilation}
\author{Yu Jia~\footnote{jiay@ihep.ac.cn}}
\affiliation{Institute of High Energy Physics, Chinese Academy of Sciences, Beijing 100049, China\vspace{0.2cm}}
\affiliation{School of Physical Sciences,
University of Chinese Academy of Sciences, Beijing 100049, China\vspace{0.2cm}}
\author{Yang Liu~\footnote{liuy01@ihep.ac.cn}}
\affiliation{Institute of High Energy Physics, Chinese Academy of Sciences, Beijing 100049, China\vspace{0.2cm}}
\affiliation{School of Physical Sciences, University of Chinese Academy of Sciences, Beijing 100049, China\vspace{0.2cm}}
\author{Junliang Lu~\footnote{lujl@ihep.ac.cn}}
\affiliation{School of Physics, Henan Normal University, Xinxiang 453007, China\vspace{0.2cm}}
\affiliation{Institute of High Energy Physics, Chinese Academy of Sciences, Beijing 100049, China\vspace{0.2cm}}
\author{Guang Tang~\footnote{tangg@ihep.ac.cn}}
\affiliation{Institute of High Energy Physics, Chinese Academy of Sciences, Beijing 100049, China\vspace{0.2cm}}
\affiliation{School of Physical Sciences, University of Chinese Academy of Sciences, Beijing 100049, China\vspace{0.2cm}}
\author{Xiaonu Xiong~\footnote{xnxiong@csu.edu.cn}}
\affiliation{School of Physics and Electronics, Central South University, Changsha 410083 , China\vspace{0.2cm}}

\date{\today}

\begin{abstract}
In this work, we investigate a novel production mechanism of vector mesons, 
exemplified by the production of a neutral vector meson associated with a lepton pair in $e^+e^-$ annihilation, 
{\it i.e.}, $e^+e^-\to V l^+l^-$ ($V=J/\psi, \rho^0, \omega, \phi$, and $l=\mu, \tau$).
These vector meson production channels can be precisely accounted within QED.
The production rates of these processes are dominated by those diagrams where the vector meson is emitted from 
either the incident electron or positron, which exhibit a $\ln^2 m_l^2$ enhancement stemming from the triple collinear limit of leptons.
Our numerical analysis indicates that the corresponding production rates are substantial enough to warrant the observation 
of these novel vector meson production channels at {\tt{BESIII}} and {\tt{Belle II}} experiments in near future.
\end{abstract}

\maketitle

\section{Introduction}

The vector meson production at $e^+e^-$ colliders provides a fruitful platform to enrich our understanding toward hadron physics.
As is well known, the neutral vector mesons, exemplified by $J/\psi$, $\rho^0$, $\omega$, and $\phi$,
can be copiously produced on the resonance peak in $e^+e^-$ annihilation.
Moreover, both inclusive and exclusive $J/\psi$ productions at high-energy $e^+e^-$
colliders such as $B$ factories,
have attracted lots of interest during the past decades, which serves as a sensitive probe into the interplay between perturbative and nonperturbative QCD. In particular, the inclusive $J/\psi$ production and exclusive
double charmonium production processes~\cite{Brambilla:2010cs}, {\it e.g.}, $e^+e^-\to J/\psi+\eta_c(\chi_{cJ}),\,J/\psi J/\psi$,
are widely viewed as the golden channels to test the validity of the nonrelativistic QCD (NRQCD) factorization approach~\cite{Bodwin:1994jh}.

The goal of this work is to investigate a new class of vector meson production channels, which have very clean signatures
and can be readily observed at {\tt Belle II} and {\tt BESIII} experiments.
Concretely speaking, we perform a comprehensive investigation on the reactions
$e^+e^-\to V l^+ l^-$ ($l=\mu,\tau$), where the neutral vector meson is produced simply via photon conversion.
These production processes are essentially QED-like, and therefore it is conceivable that future experimental measurement
should match the theoretical predictions perfectly.

On the phenomenological ground, precise knowledge of $e^+e^-\to V l^+ l^-$ is also important. For example,
the reaction $e^+e^-\to J/\psi \mu^+ \mu^-$ constitutes the major background for the long-sought
$e^+e^-\to J/\psi J/\psi$ signals.
Another merit of these channels, especially at {\tt BESIII} energy, is that the
$V l^+ l^-$ signals are entirely produced from the $e^+e^-$ continuum,
and one does not need to worry about the contamination from the electromagnetic transition such as
$\psi(3770)\to \gamma^*l^+l^-\to V l^+l^-$, which is forbidden by $C$ parity conservation.

\section{Numerical results}
\label{sec:numerical}

In the processes considered in this work, the neutral vector meson $V$ is produced purely electromagnetically, {\it i.e.}, via the $\gamma\to V$ conversion. Although this production mechanism is often
referred to as the {\it vector meson dominance} (VMD) model, we emphasize that the calculational framework presented here is
model independent.

The decay constant of the neutral vector meson is defined by
\beq
\langle V(P,\lambda) \vert J_{\mathrm{EM}}^\mu \vert 0 \rangle =  g_V\,\varepsilon^{*\mu}(P,\lambda),
\label{gv:def:decay:constant}
\eeq
where $J_{\mathrm{EM}}^\mu$ denotes the electromagnetic current and $\varepsilon^{*\mu}$ represents the
polarization vector. The values of $g_V$ of different neutral vector mesons can be accurately
determined through the measured leptonic width of $V\to e^+e^-$, whose values are enumerated in
Table~\ref{Table:gV:four:vector:mesons}.

\begin{table}[hbtp]
\centering
  \begin{tabular}{|c|c|c|c|c|}
     \hline
 $V$            &$J/\psi$         &$\rho^0$          &$\omega$        &$\phi$ \\  \hline
 $\;M_V~[\mathrm{GeV}]\;$  &$\;3.0969\;$ &$\;0.7753\;$   &$\;0.7827\;$    &$\;1.0194\;$ \\ \hline
$\;g_V~[\mathrm{GeV}^2]\;$ &$\;0.8319\;$   &$\;0.1177\;$   &$\;0.0360\;$   &$\;0.0753\;$\\
\hline
\end{tabular}
\caption{Masses and $g_V$ for various neutral vector mesons.}
\label{Table:gV:four:vector:mesons}
\end{table}

\begin{figure}[hbt]
\begin{center}
\includegraphics[width=0.5\textwidth]{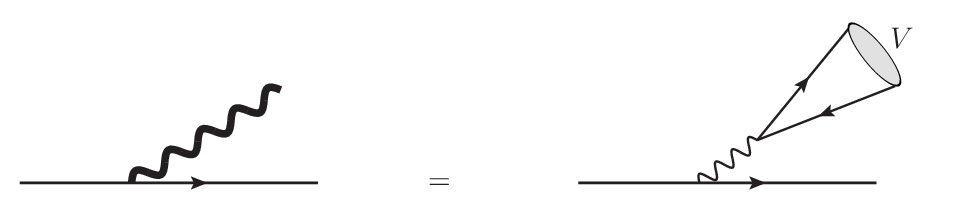}
\caption{An effective $llV$ vertex constructed from the VMD.}
\label{llV:vertex:Feynman:Rule}
\end{center}
\end{figure}

As indicated in Fig.~\ref{llV:vertex:Feynman:Rule}, one may facilitate
the calculation by assuming that the neutral vector meson is directly radiated off the lepton line,
with the effective $llV$ coupling represented by $(4\pi\alpha g_V/M^2_V)\gamma^\mu$.

\begin{figure}[hbt]
\begin{center}
\includegraphics[width=0.6\textwidth]{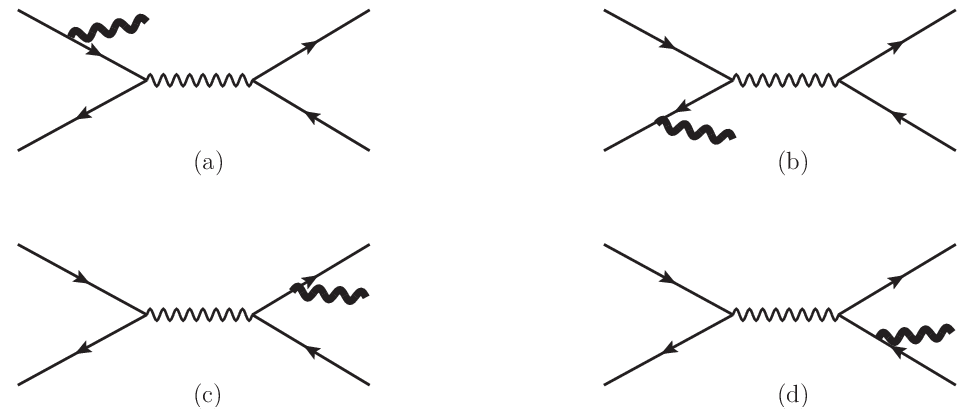}
\caption{Four lowest-order diagrams for $e^+e^-\to V+ l^+l^-$ ($l=\mu,\tau$).}
\label{LO:Feynman:Diagrams}
\end{center}
\end{figure}

There are four lowest-order (LO) QED diagrams for $e^+e^-\to V l^+l^-$ ($l=\mu,\tau$), as shown in Fig.~\ref{LO:Feynman:Diagrams}.
It is interesting to note that these diagrams can be generated by crossing the corresponding diagrams for
the rare four lepton decay of neutral vector meson exemplified by $J/\psi\to \mu^+\mu^-+e^+e^-$~\cite{Chen:2020bju}.
It is useful to divide these four diagrams into two classes, depending on which lepton $V$ is emitted from.
The first two diagrams in Fig.~\ref{LO:Feynman:Diagrams} are dubbed the initial-state emission (ISE) processes, while the last two
are referred to as the final-state emission (FSE) processes.

Our goal in this work is to assess the observation potential of the $V l^+l^-$ production processes at two $e^+e^-$ colliders in commission: {\tt Belle II} with $\sqrt{s}=10.58$ GeV and {\tt BESIII} with $\sqrt{s}=3.77$ GeV.
The masses of various vector mesons, $m_\mu=0.10566\;\mathrm{GeV}$ and $m_\tau=1.77693\;\mathrm{GeV}$ are taken from the latest compilation of the Particle Data Group~\cite{ParticleDataGroup:2024cfk}. We choose the running QED coupling $\alpha(10.58\;\mathrm{GeV})=1/131.0$ and $\alpha(3.77\;\mathrm{GeV})=1/132.5$~\cite{Bierlich:2022pfr}.
The numerical integration is conducted with the aid of the package {\tt{HCubature}}~\cite{Hahn:2004fe}.

\begin{table}[hbtp]
    \centering
    \begin{tabular}{|c|c|l|c|c|c|c|c|c|c|c|}
      \multicolumn{11}{c}{$e^+e^-\to V\mu^+\mu^-$}\\
      \hline
 & \multicolumn{5}{|c|}{{\tt {BESIII}}~[fb]}& \multicolumn{5}{|c|}{{\tt{Belle II}}~[fb]}\\\hline\hline
 $V$& \multicolumn{2}{|c|}{$\sigma_{\rm{ISE}}$}& $\sigma_{\rm{FSE}}$& \multicolumn{2}{|c|}{$\sigma_{\rm{TOT}}$}& \multicolumn{2}{|c|}{$\sigma_{\rm{ISE}}$}& $\sigma_{\rm{FSE}}$& \multicolumn{2}{|c|}{$\sigma_{\rm{TOT}}$}\\\hline \hline
         $J/\psi$&   \multicolumn{2}{|c|}{\,\,227.905\,\,}&  \,\,0.002\,\,&  \multicolumn{2}{|c|}{\,\,227.907\,\,}&\multicolumn{2}{|c|}{\,\,\,\,54.664\,\,\,\,}&  \,\,0.907\,\,&  \multicolumn{2}{|c|}{\,\,\,\,55.571\,\,\,\,}\\ \hline
         $\rho^0$&   \multicolumn{2}{|c|}{1476.460}&  82.156&  \multicolumn{2}{|c|}{1558.620}&   \multicolumn{2}{|c|}{479.461}&  49.080&  \multicolumn{2}{|c|}{505.541}\\ \hline
         $\omega$&   \multicolumn{2}{|c|}{131.986}&  7.227&  \multicolumn{2}{|c|}{139.213}&   \multicolumn{2}{|c|}{40.856}&  4.361&  \multicolumn{2}{|c|}{45.217}\\ \hline
         $\phi$&   \multicolumn{2}{|c|}{177.555}&  5.950&  \multicolumn{2}{|c|}{183.505}&   \multicolumn{2}{|c|}{56.370}&  4.901&  \multicolumn{2}{|c|}{61.271}\\ \hline
    \end{tabular}
    \quad
    \begin{tabular}{|c|c|c|c|c|c|}
      \multicolumn{6}{c}{$e^+e^-\to V\tau^+\tau^-$}\\\hline
      & \multicolumn{5}{|c|}{{\tt{Belle II}}~[fb]}\\\hline\hline
      $V$& \multicolumn{2}{|c|}{$\sigma_{\rm{ISE}}$}& $\sigma_{\rm{FSE}}$& \multicolumn{2}{|c|}{$\sigma_{\rm{TOT}}$} \\\hline \hline
              $J/\psi$&   \multicolumn{2}{|c|}{\,\,\,\,6.055\,\,\,\,}&  \,\,0.365\,\,&  \multicolumn{2}{|c|}{\,\,6.420\,\,} \\ \hline
              $\rho^0$&   \multicolumn{2}{|c|}{118.887}&  24.609&  \multicolumn{2}{|c|}{143.496} \\ \hline
              $\omega$&   \multicolumn{2}{|c|}{10.603}& 2.188&  \multicolumn{2}{|c|}{12.791} \\\hline
              $\phi$&   \multicolumn{2}{|c|}{13.229}&  2.477&  \multicolumn{2}{|c|}{15.706} \\\hline
         \end{tabular}
   \caption{Predictions for $\sigma(e^+e^-\to V+l^+l^-)$ with $V=J/\psi,\rho^0,\omega,\phi$ and $l=\mu, \tau$ at {\tt Belle} and {\tt BESIII}
   center-of-mass energies. We also enumerate the individual contribution from the ISE and FSE channels.
   It happens that the interference between the ISE and FSE amplitudes yields a vanishing contribution to the integrated cross section,
   so we do not list their values here. }
\label{Table:Integrated:X:Sections:ISE:FSE}
\end{table}

In Table~\ref{Table:Integrated:X:Sections:ISE:FSE} we present the numerical predictions for $\sigma(e^+e^-\to V+l^+l^-)$
at {\tt Belle} and {\tt BESIII} energies, with $V=J/\psi,\rho^0,\omega,\phi$, and $l=\mu, \tau$.
The magnitudes of the production rates associated with various channels appear to be sizable.
We can estimate the yields of the signal events at both experiments.
Up to now the {\tt{BESIII}} and {\tt{Belle II}} experiments have already achieved about $20\;\mathrm{fb}^{-1}$ and $1500\;\mathrm{fb}^{-1}$ integrated luminosity.
Table~\ref{Table:Integrated:X:Sections:ISE:FSE} indicates that the numbers of the signal events for $J/\psi$, $\rho^0$, $\omega$, $\phi$
associated with a $\mu^+\mu^-$ pair are $4458$, $31172$, $2784$, $3670$ at {\tt{BESIII}},
and are $83357$, $758311$, $67825$, $91907$  at {\tt{Belle II}}, respectively.
There is no enough phase space for {\tt BESIII} to observe the $V+\tau^+\tau^-$ channels, and the yields of the
corresponding signal events are anticipated to be $9630$, $215243$, $19186$, $23560$ at {\tt Belle II}.

\begin{figure}[htb]
\includegraphics[width=0.6\textwidth]{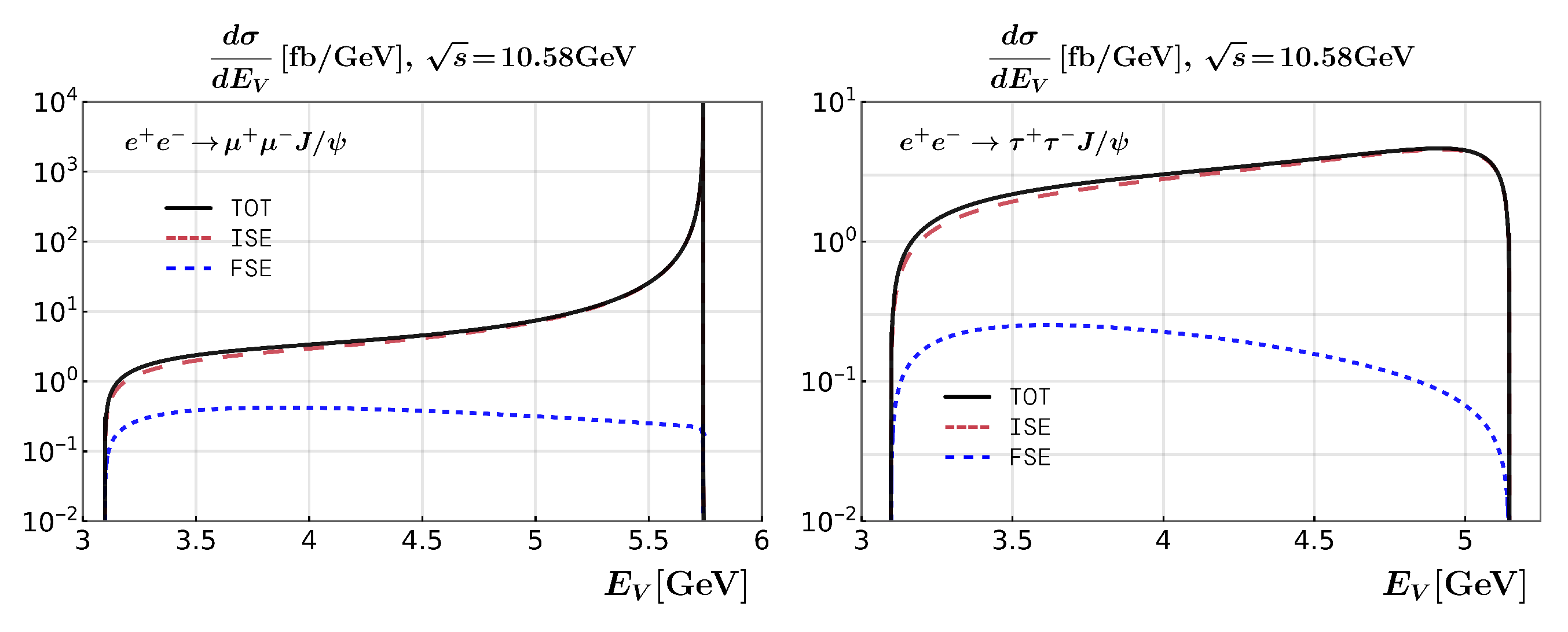}
\includegraphics[width=0.6\textwidth]{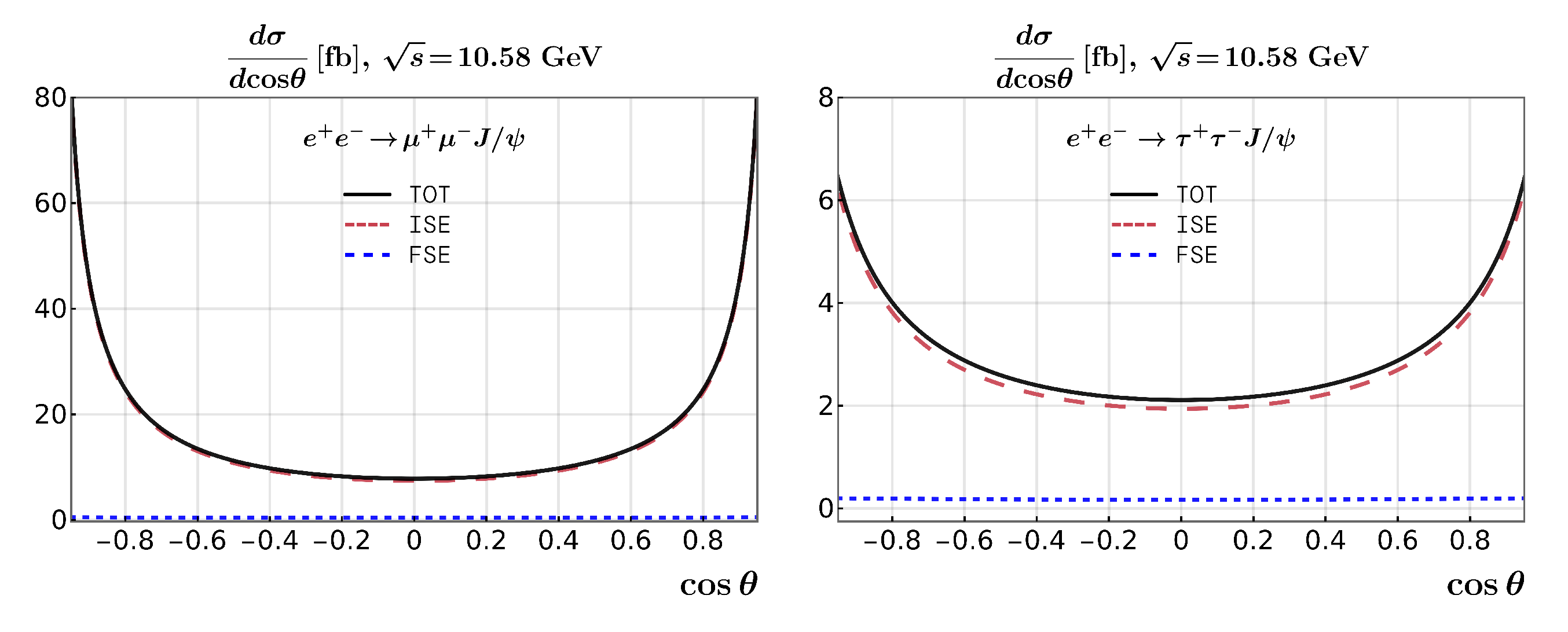}
 \caption{Energy and angular distributions of $J/\psi$ in $e^+e^- \to J/\psi+\mu^+\mu^-(\tau^+\tau^-)$ at {\tt Belle II} energy in the center-of-mass frame.}
 \label{Jpsi:dimuon:distributions:BELLE}
\end{figure}

\begin{figure}[htb]
\includegraphics[width=0.6\textwidth]{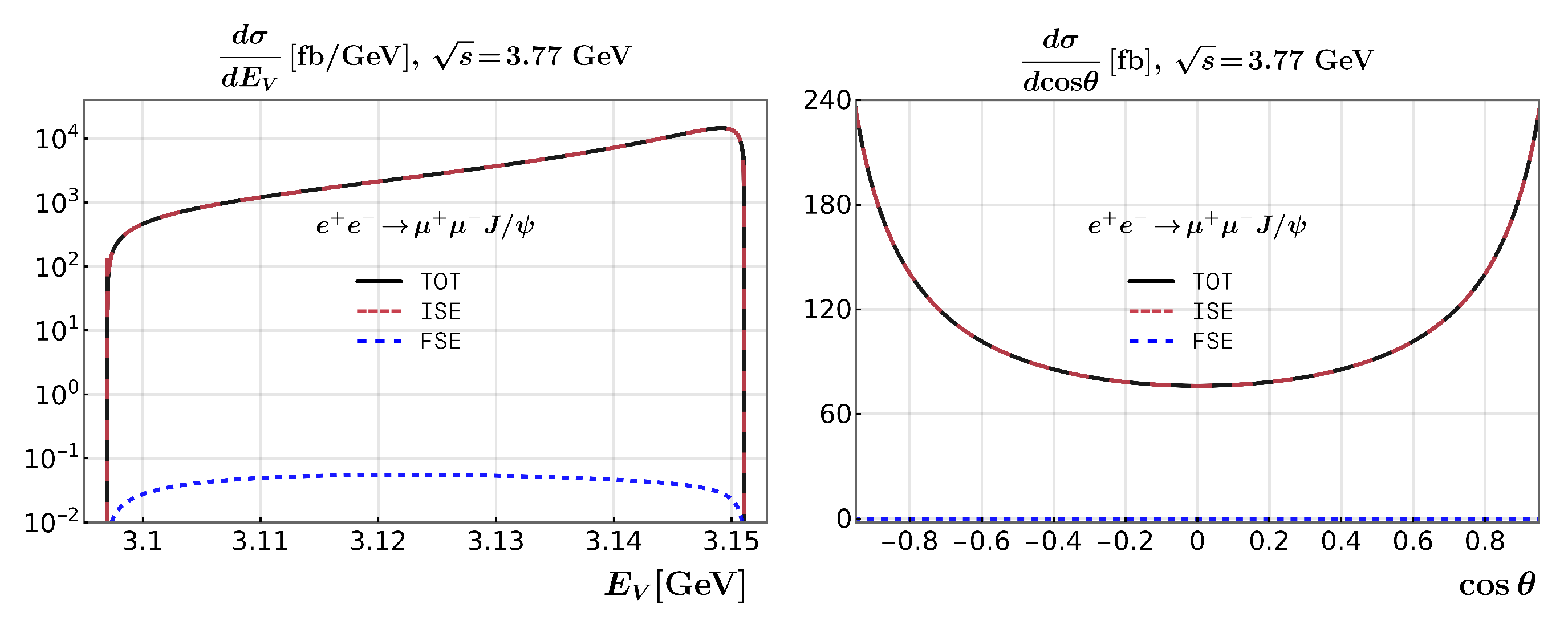}
\caption{Energy and angular distributions of $J/\psi$ in $e^+e^- \to J/\psi+\mu^+\mu^-$
at {\tt BESIII} energy in the center-of-mass frame.}
\label{Jpsi:dimuon:distributions:BESIII}
\end{figure}

It is interesting to compare $e^+e^-\to V l^+l^-$ with similar reactions, $e^+e^-\to \gamma^*\gamma^* \to V_1 V_2$, which are largely mediated by the double VMD mechanism~\cite{Davier:2006fu, Bodwin:2006yd}.
Despite the suppression brought by the three-body phase space, the production rates of the former
are still considerably greater than those of the latter. It is worth pointing out that,
the reaction $e^+e^-\to J/\psi+\mu^+\mu^-$ constitutes one of the dominant backgrounds for the long-sought exclusive double charmonium production
reaction $e^+e^-\to J/\psi J/\psi$, whose cross section is estimated to be around $2.13^{+0.30}_{-0.06}$ fb according to
the state-of-the-art NRQCD prediction~\cite{Sang:2023liy, Huang:2023pmn}.

Taking $e^+e^-\to J/\psi+\mu^+\mu^-$ as a benchmark reaction,
we  in Fig.~\ref{Jpsi:dimuon:distributions:BELLE} and Fig.~\ref{Jpsi:dimuon:distributions:BESIII} have plotted
$J/\psi$'s energy and angular distributions at {\tt Belle 2} and {\tt BESIII} experiments, respectively.
The dominant contributions to the cross sections appear to come from the regions
$E_{J/\psi}\approx (s+M_{J/\psi}^2-4m_\mu^2)/2\sqrt{s}$ and $\theta \approx 0,\, \pi$.

\begin{figure}[htb]
\includegraphics[width=0.4\textwidth]{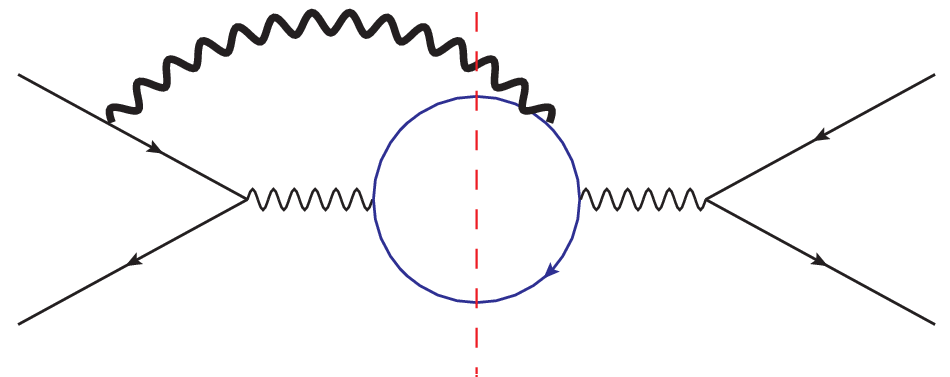}
 \caption{A representative cut diagram by stitching one ISE diagram
 with another FSE diagram in Fig.~\ref{LO:Feynman:Diagrams}, which is equivalent to the interference
 contribution to the cross section. }
 \label{Cut:diagram:Interference:ISE:FSE}
\end{figure}

In Table~\ref{Table:Integrated:X:Sections:ISE:FSE} and Fig.~\ref{Jpsi:dimuon:distributions:BELLE} and Fig.~\ref{Jpsi:dimuon:distributions:BESIII}, we also
list the individual contributions from the ISE and FSE channels.
An important observation is that, the ISE contribution dominates over the FSE for virtually
all the reactions. We devote the next section to a comprehensive theoretical explanation.

It is of significance to explain why the interference contribution is absent in Table~\ref{Table:Integrated:X:Sections:ISE:FSE}.
A typical contribution from the interference term is illustrated by the cut diagram in Fig.~\ref{Cut:diagram:Interference:ISE:FSE},
in which three neutral vector particles (with $C=-1$) are attached to the muon loop. Guaranteed by Furry's theorem,
the interference between ISE and FSE yields a vanishing contribution to the integrated cross section (Actually the interference contribution is also absent in the differential energy and angular distributions of $J/\psi$ in Fig.~\ref{Jpsi:dimuon:distributions:BELLE} and Fig.~\ref{Jpsi:dimuon:distributions:BESIII}).

\section{Asymptotic behavior of the cross section}
\label{Asymp:Behav:X:Section}

The reactions $e^+e^-\to V+ l^+l^-$ ($l=\mu,\tau$) entail multiple energy scales.
It is safe to neglect the electron mass at the outset, consequently
the integrated cross section involves three remaining scales: $m_l$, $M_V$ and $\sqrt{s}$.
Taking $l=\mu$ for concreteness,  the most natural working assumption is then
$m_l\ll \sqrt{s}$ and $m_l \ll M_V$. In this section, we deduce the asymptotic behaviors of the cross sections from
the ISE and FSE channels, and understand the reason why ISE dominates over FSE.

\subsection{Asymptotic behavior of the ISE contribution}

Upon squaring the ISE amplitude for $e^+e^-\to Vl^+l^-$, which is represented by the
first two diagrams in Fig.~\ref{LO:Feynman:Diagrams}, we can safely set $m_l=0$.
It is straightforward to conduct threefold phase space integration analytically and end up with
the final integration with respect to the squared invariant mass of $l^+l^-$:
\bqa
\sigma_{\mathrm{ISE}} &= & \int_{4m_l^2}^{s+M_V^2-2M_V\sqrt{s}} d\sll\frac{16 \pi \alpha^4 g_V^2}{3s^2M_V^4  \sll (s-M_V^2-\sll)} \Bigg[\left(s^2+M_V^4+2 M_V^2 \sll+\sll^2\right)
\nn\\
&  \times& \ln \frac{s-\sll-M_V^2+\sqrt{\lambda(s,\sll,M_V^2)}}{s-\sll-M_V^2-\sqrt{\lambda(s,\sll,M_V^2)}} -2 \left(s-M_V^2-\sll\right) \sqrt{\lambda(s,\sll,M_V^2)}\Bigg],
\label{sll:inv:distribution:ISE:X:Section}
\eqa
with $\lambda(x,y,z)\equiv x^2+y^2+z^2-2xy-2xz-2yz$ signifying the K\"{a}llen function.
Notice that the lepton mass enters in the lower bound of the integral.
It is infeasible to carry out this integration directly.

As is evident in the energy spectra shown in Fig.~\ref{Jpsi:dimuon:distributions:BELLE} and Fig.~\ref{Jpsi:dimuon:distributions:BESIII},
the dominant contribution to the ISE cross section
comes from the region where $E_V$ approaches its maximum, or equivalently, $\sll \to 0$.
Therefore it becomes legitimate to expand the integrand in Eq.~\eqref{sll:inv:distribution:ISE:X:Section} in powers of $\sll$,
and then conduct the integration at leading order,
\bqa
\sigma_{\mathrm{ISE}} &=& \frac{8\pi\alpha^{4}g_{V}^{2}(s^{2}+M_V^{4})}{3s^{2}M_V^{4}(s-M_V^{2})}
    \Bigg[\ln{\frac{\Lambda^2 }{4m_{l}^2}}\ln{\frac{s^2}{4m_{l}^2\Lambda^{2}}}+2\ln{\frac{\Lambda^2}{4m_{l}^2}}\ln{\frac{(s-M_V^2)^2}{s M_V^2}}\Bigg]
\nn \\
& -& \frac{32\pi\alpha^{4}g_V^2(s-M_{V}^{2})}{3s^{2}M_{V}^{4}} \ln{\frac{\Lambda^{2}}{4m_{l}^{2}}}+ {\cal O}(m_l^0),
\label{ISE:ASYM::FORM}
\eqa
where $\Lambda^2$ specifies the upper bound of the integral.
It is legitimate to take $\Lambda\sim {\cal O}(\sqrt{s})$ or ${\cal O}(M_V)$,
whose accurate value becomes irrelevant.

The key message conveyed in Eq.~\eqref{ISE:ASYM::FORM} is that,
the ISE contribution is dominated by the double logarithmic factor $\ln^2{m_l^2}$.
This double logarithmic enhancement actually originates from the triple collinear region where both the $l^+$ and
$l^-$ fly nearly parallel to the incident $e^-$ or $e^+$ beam direction.
In Appendix~\ref{app:asympt}, we provide an alternative approach to reproduce this asymptotic $\ln m_l^2$ behavior,
based on the factorization formula entailing the QED fracture function.

\subsection{Asymptotic behavior of the FSE contribution}

We then turn to the last two diagrams in Fig.~\ref{LO:Feynman:Diagrams}. Since
the lepton propagator is always far off-shell in the FSE channel, it is a good approximation to set $m_l=0$.
Conducting the three-body phase space integration analytically, we obtain the following integrated cross section:
\bqa
\sigma_{\mathrm{FSE}} &= & \sigma (e^{+}e^{-}\to l^{+}l^{-})\frac{2g_V^2\alpha^2}{M_V^4} \Bigg\{(1+r)^2\ln{\frac{1}{r}}\ln{\frac{(1+r)^{4}}{r}}-(3 + 4r + 3r^2 )\ln{\frac{1}{r}}
\nn\\
&- & \frac{1}{3}(1+r)\Big[r(\pi^{2}+15) + (\pi^{2}-15)\Big]-4(1 + r)^{2}\mathrm{Li}_{2}\left(-r\right)\Bigg\},
\label{Exact:FSE:X:Section}
\eqa
where $\sigma(e^{+}e^{-}\to l^{+}l^{-})=(4\pi \alpha^2)/(3s)$, and $r \equiv M_V^2/s$.

In the limit $\sqrt{s}\gg M_V$, Eq.~\eqref{Exact:FSE:X:Section} reduces to
\beq
\sigma_{\mathrm{FSE}}  \approx \hat{\sigma}(e^{+}e^{-}\to l^{+}l^{-})\frac{2g_V^2\alpha^2}{M_V^4} \Bigg(\ln^2{\frac{1}{r}} -3\ln{\frac{1}{r}}+5-\frac{\pi^{2}}{3}\Bigg),
\label{FSE:ASYM::FORM}
\eeq
with the leading asymptotical behavior characterized by the Sudakov double logarithm.
The occurrence of the $\ln^2 r$ term comes from the configuration where $V$ becomes
simultaneously collinear and soft.

The logarithmic pattern in Eq.~\eqref{FSE:ASYM::FORM} can also be understood through the fragmentation approximation.
Following the fragmentation treatment of $Z^0\to J/\psi l^+l^-$~\cite{Fleming:1994iu}, we can express
the differential cross section for $e^{+}e^{-}\to V l^{+}l^{-}$ from the FSE processes as
\bqa
{d\sigma_\mathrm{Frag}\left(e^{+}e^{-}\to V(E_{V})+l^{+}l^{-}\right)\over dE_{V}}\bigg\vert_{\rm FSE} &=&
\frac{4}{\sqrt{s}}\hat{\sigma}(e^{+}e^{-}\to l^{+}l^{-})D_{l \to V }\left(\frac{2E_{V}}{\sqrt{s}},\mu^{2}\right)
\nn \\
&+ & \frac{d\hat{\sigma}}{dE_{\gamma}}\big(e^{+}e^{-}\to \gamma(E_{\gamma})+l^{+}l^{-},\mu^{2}\big)P_{\gamma\to V},
\label{Fragmentation:Approximation:FSE}
\eqa
where $D_{l\to V}(z)$ denotes the lepton-to-$V$ fragmentation function, $P_{\gamma/V}$ signifies the
probability of a photon converting into $V$, and $\mu$ is the factorization scale.
Integrating  Eq.~\eqref{Fragmentation:Approximation:FSE} with respect to $E_V$, we then recover
both the double and single logarithms in Eq.~\eqref{FSE:ASYM::FORM}.

To conclude, both ISE and FSE contributions are dominated by two different kinds of double logarithms.
The reason that the ISE contribution overwhelms the FSE contribution
is largely due to the fact $\ln^2 m_l^2\gg \ln^2 r$.

\section{summary}

In this work we explore a new production mechanism of vector meson at electron-positron colliders.
We have investigated the associated production of a neutral vector meson with a lepton pair in $e^+e^-$ annihilation,
{\it i.e.}, $e^+e^-\to V l^+l^-$ ($V=J/\psi, \rho^0, \omega,\phi$ and $l=\mu,\;\tau$).
In the context of vector meson dominance, these vector meson production channels can be accurately described within QED. 
Through numerical and analytic study, 
it is found that the production cross sections of these channels are dominated by 
those sub-diagrams referred to as the ISE type, 
{\it viz.}, where the vector meson is radiated off from the incoming electron and positron.
The leading asymptotical behavior of the ISE contribution is dictated by the 
$\ln^2 m_l^2$ term, which has to originate from the triple collinear limit of leptons. 
We have provided a novel route to reproducing this leading double logarithm, from the perspective of the
QED factorization framework that entails lepton fracture function. 
The numerical analysis indicates that the production rates of numerous 
vector meson production channels are quite sizable, considerably larger than those of $e^+e^-\to \gamma^*\gamma^*\to VV$ 
processes.
The observation prospects of these 
novel vector meson production channels appear very bright at {\tt{Belle II}} and {\tt{BESIII}} 
experiments.

\begin{acknowledgments}
We thank Jingshu Dai, Siwei Hu, Suxian Li and Chengping Shen for useful discussions.
The work of Y.~J., Y.~L., J.~L., G.~T. is supported in part by the National Natural Science Foundation of China
under Grants No.~11925506.
The work of X.~X. is is supported in part by the National Natural Science Foundation of China
under Grants No.~12275364.
\end{acknowledgments}

\appendix
\section{Asymptotic behavior of the ISE contribution from the fracture function approach}
\label{app:asympt}

\begin{figure}[htb]
    \centering
\includegraphics[width=0.35\textwidth]{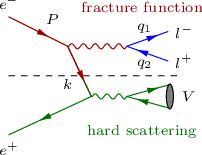}
    \caption{A schematic illustration of the factorization program valid in the triple collinear limit of the ISE channel,
    where both $l^+$ and $l^-$ move nearly parallel to the electron beam direction.}
\label{Illustration:QED:Factorization}
\end{figure}

In this appendix, we employ an alternative approach to reproduce the asymptotic $\ln m_l^2$ scaling in Eq.~\eqref{ISE:ASYM::FORM},
based on the factorization ansatz.
As noted earlier, the leading contribution from the ISE channel comes from the triple collinear limit,
where $l^+$, $l^-$ move almost collinear to the $e^-$ beam direction.
In such a limit, the virtuality of the electron propagator in Fig.~\ref{Illustration:QED:Factorization} becomes small.
It is persuasive to treat this virtual electron as an active parton, which in turn participates in the hard scattering process
$e^- e^+\to V$.
Intuitively, one may anticipate that the differential ISE cross section can be written as the convolution
of the partonic cross section with some universal long-distance function,
which characterizes the probability to find an electron parton carrying a certain momentum fraction, accompanied
by the tagged $l^+$ and $l^-$ which move in the forward direction.

As a matter of fact, such a universal long-distance function has been introduced in QCD long ago,
as a generalization of the familiar parton distribution function (PDF).
The {\it fracture function}, originally introduced by Trentadue and Veneziano~\cite{Trentadue:1993ka},
characterizes the probability of finding a parton inside a hadron accompanied by other identified hadrons in the beam remnants.
In some sense, the fraction function is an admixture of the PDF and fragmentation function.
To date our knowledge about nucleon's fracture function is rather primitive,
largely due to the lack of experimental data as well as the impossibility of making predictions
from lattice QCD.

Fortunately, the fracture function we are interested in is framed entirely in QED,
which is thus amenable to perturbation theory.
We take the incident electron to move along the $z$ direction.
It is convenient to adopt the light-cone coordinates $(p^+,\boldsymbol{p}_\perp,p^-)$ with $p^\pm=(p^0\pm p^3)/\sqrt{2}$.
The QED fraction function related to $e\to e(x)+l^+l^-$ can be
defined as the following light-cone correlator~\cite{Collins:2011zzd}:
\beq
    \begin{aligned}
        M^{l^+l^-}_{e/e}(\{x; y_1,\boldsymbol{q}_{1\perp}; y_2,\boldsymbol{q}_{2\perp}\})&=\int\frac{dw^-}{4\pi}e^{-ixP^+w^-}\langle e(P)|\bar{\psi} (0,w^-,\boldsymbol{0}_\perp)\mathcal{W}(\infty,w^-)^\dagger|l^+(q_1)l^-(q_2)\rangle\\
        &\times \langle l^+(q_1)l^-(q_2) \left|\gamma^+\mathcal{W}(\infty,0)\psi(0) \right| e(P)\rangle,
    \end{aligned}
\eeq
where $x$, $y_{1,2}$ signify the light-cone momentum fraction carried by the electron parton and by
the beam remnants $l^\mp$.  $\cal{W}$ represents a light-like Wilson line, whose role is to ensure QED gauge invariance.

\begin{figure}[htb]
    \centering
    \includegraphics[width=0.4\textwidth]{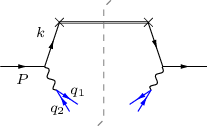}
    \caption{Leading-order diagram for the QED fracture function associated with $e\to e+l^+l^-$.}
    \label{fig:feynman:frac}
\end{figure}

It is convenient to work with the light-cone gauge $A^+ = 0$, so that the gauge link can be dropped.
At lowest order in QED, there is only one Feynman diagram for the fracture function related to $e\to e(x)+l^+l^-$,
which is shown in Fig.~\ref{fig:feynman:frac}.
Neglecting all the lepton masses, the intended QED fracture function reads
\beq
    \begin{aligned}
      M^{l^+l^-}_{e/e}(x;& y_1, \boldsymbol{q}_{1\perp}; y_2, \boldsymbol{q}_{2\perp}) = \frac{e^4}{4}\int d^4k \ \delta(k^+ - xP^+)\delta^{(4)}(P-k-q_1-q_2)\\
        &\times \sum_{\text{spins}}\bar{u}(P)\gamma^\mu\frac{i}{\slashed{k}+i\epsilon}\bar{v}(q_2)\gamma^\nu u(q_1)\gamma^+\bar{u}(q_1)\gamma^\rho v(q_2)\frac{i}{\slashed{k}+i\epsilon}\gamma^\sigma u(P)\\
        &\times D_{\mu\nu}(q) D_{\rho\sigma}(q),
    \end{aligned}
\label{Fracture:function:expression}
\eeq
where $D_{\mu\nu}$ denotes the photon propagator in light-cone gauge:
\begin{equation}
    D_{\mu\nu}(q)=\frac{-i}{q^2 + i\epsilon} \left(g_{\mu\nu} - \frac{q_\mu n_\nu + q_\nu n_\mu}{n\cdot q}\right),
\end{equation}
with $n^\mu$ being a reference null vector such that $q\cdot n=q^+$.

It is straightforward to work out the integration in Eq.~\eqref{Fracture:function:expression}, and
arrive at
\begin{equation}
    \begin{aligned}
        M^{l^+l^-}_{e/e}(&x; y_1, \boldsymbol{q}_{1\perp}; y_2, \boldsymbol{q}_{2\perp}) = 4e^4 \ \delta(1-x-y_1-y_2) y_1^3 y_2^3  \\
        &\times \Big\{\boldsymbol{q}_{1\perp}^2\boldsymbol{q}_{2\perp}^2\big[\frac{y_1^2}{2} (1+x^2-y_2(2-6x-4y_2))+y_1y_2(x(4x+3y_2)-y_2)+\frac{y_2^2}{2}(1+x^2)\big] \\
        &+ \boldsymbol{q}_{1\perp}^4 y_2^2 \big[1 - x(4-5x-6y_1) - 2 y_1(1-y_1)\big]-4(\boldsymbol{q}_{1\perp}\cdot\boldsymbol{q}_{2\perp})^2 y_1 y_2^2 (2x + y_2) \\
        &- 2 (\boldsymbol{q}_{1\perp}\cdot\boldsymbol{q}_{2\perp})\big[\boldsymbol{q}_{1\perp}^2y_2 (x^2 (2 y_1+y_2)+x (y_1^2-5y_1 y_2 +2 y_2)+y_1^2-y_2(1-3 y_1+4 y_1^2) )\big]\Big\}\Big/\ \\
        &\Big\{(1-x)^2(\boldsymbol{q}_{1\perp}y_2-\boldsymbol{q}_{2\perp}y_1)^4\big[\boldsymbol{q}_{1\perp}^2y_2+\boldsymbol{q}_{2\perp}^2y_1-(\boldsymbol{q}_{1\perp}y_2-\boldsymbol{q}_{2\perp}y_1)^2\big]^2\Big\} + (y_1\leftrightarrow y_2, \boldsymbol{q}_{1\perp}\leftrightarrow\boldsymbol{q}_{2\perp}).
    \end{aligned}
\label{Fracture:Function:Closed:Form}
\end{equation}

Eq.~\eqref{Fracture:Function:Closed:Form} is an interesting byproduct of this work. We stress that this fracture function is universal.
It might be utilized in other processes, {\it e.g.}, $e^+e^-\to Z^0\mu^+\mu^-$ at future {\tt CEPC} and {\tt FCC} experiments.

According to the factorization formula picturized in Fig.~\ref{Illustration:QED:Factorization},
we can express the total ISE cross section as the following convolution integral:
\beq
\sigma_{\mathrm{ISE}}=2 \int_0^1dx\int\frac{d^2\boldsymbol{q}_{1\perp}}{(2\pi)^3} \frac{d^2\boldsymbol{q}_{2\perp}}{(2\pi)^3}\frac{dy_1}{2y_1}\frac{dy_2}{2y_2}M(x, y_1, y_2, \boldsymbol{q}_{1\perp}, \boldsymbol{q}_{2\perp})\hat{\sigma}(e^+e^-\to V),
\label{ISE:X:Section:Factorization:Fracture}
\eeq
where the partonic cross section is
\beq
\hat{\sigma}(e^+e^-\to V)=\frac{2\pi e^4g_V^2}{sM_V^4}\delta\left(x-\frac{M_V^2}{s}\right).
\label{eq:hard:kernel}
\eeq
The prefactor 2 in Eq.~\eqref{ISE:X:Section:Factorization:Fracture} simply
reflects the fact that the fracture function of
$e^+\to e^++l^+l^-$ is identical to that of $e^-\to e^-+l^+l^-$ by charge conjugation symmetry,
which makes the equal contribution to the total ISE cross section.

It is useful to insert
the following identity,
\begin{equation}
    \begin{aligned}
        \notag 1 = &\int d\sll \ \delta(\sll-(q_1+q_2)^2)\\
        =&\int d\sll \ \delta\left(2\boldsymbol{q}_{1\perp}\cdot\boldsymbol{q}_{2\perp}-\frac{y_2^2\boldsymbol{q}_{1\perp}^2+y_1^2\boldsymbol{q}_{2\perp}^2}{y_1 y_2}+\sll\right)
    \end{aligned}
    \label{eq:sll}
\end{equation}
into Eq.~\eqref{ISE:X:Section:Factorization:Fracture}, so that $s_{ll}$ can be introduced as an integration variable.

The integration of the above $\delta$-function in Eq.~\eqref{ISE:X:Section:Factorization:Fracture} not only eliminates
the azimuthal variable $\boldsymbol {q}_{1\perp}\cdot \boldsymbol {q}_{2\perp}$, but constrains
the integration interval of $\boldsymbol{q}_{1\perp}^2$.
Integrating out $\boldsymbol{q}_{2\perp}^2$ would lead to logarithmic UV divergence.
Nevertheless, this UV divergence is of no much concern since the validity of fracture function factorization
demands that the transverse momenta of the identified beam remnants, $l^+$ and $l^-$,
cannot be arbitrarily large.
We choose to impose an UV cutoff $\mu^2$ on the integration over $\boldsymbol{q}_{2\perp}^2$.
After integrating out $y_{2}$ and $\boldsymbol{q}_{1,2\perp}^2$, expanding the integrand in powers of $\sll/\mu^2$,
then integrating out $y_{1}$, we arrive at
\beq
\sigma_\mathrm{ISE} = {8\pi\alpha^4 g^2_V\over 9 M_V^4 s^2 (s-M_V^2)} \int_{4m_l^2}^{\Lambda^2} \!
{d \sll \over \sll} \, \left[6(s^2+M_V^4)\left(\ln{\frac{s}{\sll}}+\ln{\frac{\mu^2}{M^2_V}}\right)+7s^2+12sM_V^2+7M_V^4\right].
\label{Fracture:Factorization:int:over:sll}
\eeq
Obviously, it is only the term $\propto {\ln\sll/\sll}$ would yield $\ln^2 m^2_l$ after integrating out $\sll$.
Note this term is finite and does not interfere with the artificial UV cutoff $\mu^2$.

To regularize the potential UV divergence, we have also imposed an upper bound $\Lambda^2$ in Eq.~\eqref{Fracture:Factorization:int:over:sll}.
Carrying out the integration over $\sll$ in Eq.~\eqref{Fracture:Factorization:int:over:sll}, we obtain the leading double logarithmic
term:
\begin{equation}
    \sigma_\mathrm{ISE} = \frac{8\pi\alpha^4g_V^2(s^2+M_V^4)}{3s^2M_V^4(s-M_V^2)}\ln{\frac{\Lambda^2}{4 m_l^2}}\ln{\frac{s^2}{4m_l^2\Lambda^2}}
    +\cdots,
\label{Doulbe:log:Fracture:approach}
\end{equation}
which exactly matches the $\ln^2 m^2_l$ term in Eq.~\eqref{ISE:ASYM::FORM}.

\end{document}